\documentclass[fleqn,10pt]{wlscirep}
\usepackage[utf8]{inputenc}
\usepackage[T1]{fontenc}
\usepackage{tabularx}
\usepackage{wrapfig}
\usepackage{algorithm}
\usepackage{algpseudocode}
\usepackage{bm}  % for stuff like \boldsymbol -> bold math
\makeatletter
\newcommand*{\rom}[1]{\expandafter\@slowromancap\romannumeral #1@}
\makeatother

\title{Segmentation of Non-Small Cell Lung Carcinomas: Introducing DRU-Net and  Multi-Lens Distortion}
\author[1,2,*]{Soroush Oskouei}
\author[3,4,5,6]{Marit Valla}
\author[3,6,7]{André Pedersen}
\author[1,8]{Erik Smistad}
\author[3,5]{Vibeke Grotnes Dale}
\author[3,4]{Maren Høibø}
\author[5]{Sissel Gyrid Freim Wahl}
\author[5]{Mats Dehli Haugum}
\author[8,9]{Thomas Langø}
\author[10,11]{Maria Paula Ramnefjell}
\author[10,11]{Lars Andreas Akslen}
\author[9,12]{Gabriel Kiss}
\author[1,2]{Hanne Sorger}

\affil[1]{Norwegian University of Science and Technology (NTNU), Department of Circulation and Medical Imaging, Trondheim, NO-7491, Norway}
\affil[2]{Levanger hospital, Nord-Trøndelag Health Trust, Clinic of Medicine, Levanger, NO-7600, Norway}
\affil[3]{Norwegian University of Science and Technology (NTNU), Department of Clinical and Molecular Medicine, Trondheim, NO-7491, Norway}
\affil[4]{St. Olavs hospital, Trondheim University Hospital, Clinic of Laboratory Medicine, Trondheim, NO-7030, Norway}
\affil[5]{St. Olavs hospital, Trondheim University Hospital, Department of Pathology, Trondheim, NO-7030, Norway}
\affil[6]{St. Olavs hospital, Trondheim University Hospital, Clinic of Surgery, Trondheim, NO-7030, Norway}
\affil[7]{Sopra Steria, Application Solutions, Trondheim, NO-7010, Norway}
\affil[8]{SINTEF Digital, Department of Health Research, Trondheim, NO-7465, Norway}
\affil[9]{St. Olavs hospital, Trondheim University Hospital, Research Department, Center for Medical equipment, Technology, and Innovation, Trondheim, NO-7491, Norway}
\affil[10]{University of Bergen, Department of Clinical Medicine Centre for Cancer Biomarkers (CCBIO), Bergen, NO-5007, Norway}
\affil[11]{Haukeland University Hospital, Department of Pathology, Bergen, NO-5020, Norway}
\affil[12]{Norwegian University of Science and Technology (NTNU), Department of Computer Science, Trondheim, NO-7491, Norway}

\affil[*]{soroush.oskouei@ntnu.no}

\keywords{Lung carcinoma, Digital pathology, Tumor segmentation, Deep learning, Data augmentation}

\begin{abstract}
Considering the increased workload in pathology laboratories today, automated tools such as artificial intelligence models can help pathologists with their tasks and ease the workload. In this paper, we are proposing a segmentation model (DRU-Net) that can provide a delineation of human non-small cell lung carcinomas and an augmentation method that can improve classification results. The proposed model is a fused combination of truncated pre-trained DenseNet201 and ResNet101V2 as a patch-wise classifier followed by a lightweight U-Net as a refinement model.
We have used two datasets (Norwegian Lung Cancer Biobank and Haukeland University Hospital lung cancer cohort) to create our proposed model. The DRU-Net model achieves an average of 0.91 Dice similarity coefficient. The proposed spatial augmentation method (multi-lens distortion) improved the network performance by 3\%. Our findings show that choosing image patches that specifically include regions of interest leads to better results for the patch-wise classifier compared to other sampling methods. The qualitative analysis showed that the DRU-Net model is generally successful in detecting the tumor. On the test set, some of the cases showed areas of false positive and false negative segmentation in the periphery, particularly in tumors with inflammatory and reactive changes.

\end{abstract}
\begin{document}

\flushbottom
\maketitle
\thispagestyle{empty}

\section*{Introduction}

Early diagnosis of lung cancer is crucial for patient survival~\cite{rami2021future}. Although physical examinations and medical imaging are included in the diagnostic work-up, tissue samples are needed to establish a cancer diagnosis. The histopathological diagnosis including analysis of tumor biomarkers influences therapeutic decisions and should therefore be assessed as accurately and as early as possible~\cite{lim2015biomarker, woodard2016lung}.

By scanning tissue sections, the resultant whole slide images (WSIs) can be assessed on a computer screen instead of with a regular microscope. Digitization of histopathological sections may increase efficiency compared to conventional microscopy assessment and it allows the use of artificial intelligence (AI) in the analysis of WSIs~\cite{hanna2019implementation}. AI methods may increase accuracy and speed of image interpretation and have the potential to reduce inter-observer variability and refine clinical decision-making~\cite{bera2019artificial, sakamoto2020narrative, niazi2019digital}. AI can be used for automated tissue classification, identification, and segmentation~\cite{kurc2020segmentation}. Correct segmentation of the tumor is a necessary step towards computer-assisted tumor analysis and lung cancer diagnosis ~\cite{ho2021deep,qaiser2019fast, zhao2023rgsb, viswanathan2022state, wang2019artificial, davri2023deep}.

When working with WSIs, the application of AI models is complicated due to the large size of the images. One approach involves separating the images into several small squares, called patches. A patch-based analysis may, however, lead to loss  of broader spatial relationships. Alternatively, the image can be down-sampled, or a hybrid strategy that combines both methods can be used to optimize the analytical balance between detailed resolution and global context. Some of the best-performing AI methods in the analysis of WSIs are deep neural networks~\cite{cheng2023computational, davri2023deep}. The state-of-the-art in image segmentation tasks is the use of complex neural network architectures such as vision transformers and InternImage~\cite{ranftl2021vision, wang2023internimage}. However, these methods require a relatively large amount of data~\cite{park2022vision}.
Transfer learning techniques may also be used to train or fine-tune pre-trained models on new data~\cite{kassani2022deep}. Patch-wise classification (PWC) or segmentation approaches may outperform direct segmentation of the tumor in a down-sampled image without dividing it into patches~\cite{lin2018scannet}.

Several models have been proposed for tumor segmentation in WSIs~\cite{zhao2023rgsb, zeng2023mamc, wang2023dhunet, pedersen2022h2g, albusayli2021simple, chelebian2024depicter, yan2022deep, deuschel2021multi, shakeri2022fhist, titoriya2022few}. Zhao~\textit{et al.} proposed a novel hybrid deep learning framework for colorectal cancer that uses a U-Net architecture. This model features innovative residual ghost blocks, which include switchable normalization and bottleneck transformers for extracting features.~\cite{zhao2023rgsb}. 

The MAMC-Net model introduced a multi-resolution attention module that utilizes pyramid inputs for broader feature information and detail capture~\cite{zeng2023mamc}. An attention mechanism refines features for segmentation, while a multi-scale convolution module integrates semantic and high-resolution details. Finally, a connected conditional random field ensures accurate segmentation by addressing discontinuities~\cite{zeng2023mamc}. The authors showcased superior performance of their model on breast cancer metastases and gastric cancer ~\cite{zeng2023mamc}.

DHU-Net combines Swin Transformer and ConvNeXt within a dual-branch hierarchical U-shaped architecture~\cite{wang2023dhunet, liu2021swin, liu2022convnet}. This method effectively fuses global and local features by processing WSI patches through parallel encoders, utilizing global-local fusion modules and skip connections for detailed feature integration~\cite{wang2023dhunet}. The Cross-scale Expand Layer aids in resolution recovery across different scales. The network was evaluated on datasets covering different tumor features and cancer types and achieved higher segmentation results than other tested methods~\cite{wang2023dhunet}.

Pedersen \textit{et al.} introduced H2G-Net, a cascaded convolutional neural network (CNN) architecture for segmenting breast cancer regions from gigapixel histopathological images~\cite{pedersen2022h2g}. It employs a patch-wise detection stage and a convolutional autoencoder for refinement, demonstrating significant improvements in tumor segmentation. The approach outperformed single-resolution methods, achieving a Dice similarity coefficient (DSC) of (0.933±0.069)~\cite{pedersen2022h2g}. Its efficiency is underscored by fast processing times and the ability to train deep neural networks without needing to store patches on disk.

One of the most significant challenges in using WSIs for tumor segmentation is still the scarcity of labeled data. The marking of tumor tissue in WSIs by pathology experts is time-consuming, and may be a bottle neck in research. The limited availability of such annotated datasets poses a significant hurdle for the development and application of supervised learning algorithms in tumor segmentation. Given this constraint, alternative computational strategies such as unsupervised or semi-supervised learning methods should be explored. Clustering emerges as a potent tool in this context, allowing for segmentation of tumor regions with little or no need for predefined labels.~\cite{albusayli2021simple, chelebian2024depicter}.

Yan \textit{et al.} presented a self-supervised learning method using contrastive learning to process WSIs for tissue clustering~\cite{yan2022deep}. This approach generates discriminative embeddings for initial clustering, refined by a silhouette-based scheme, and extracts features using a multi-scale encoder~\cite{yan2022deep}. It achieved high accuracy in identifying tissues without annotations. Their results show an area under the curve (AUC) of 0.99 and accuracy of approximately 0.93 for distinguishing benign from malignant polyps in a cohort of 20 patients~\cite{yan2022deep}.

Few-shot learning presents a promising way of handling scarcity of labeled data~\cite{deuschel2021multi, shakeri2022fhist}. By design, few-shot learning algorithms can learn from a very limited number of labeled examples. This capability is particularly relevant for the classification of small patches, where a small set of labeled examples can guide the learning process. Few-shot learning techniques can generalize from these examples to classify new, unseen patches, facilitating the identification and segmentation of tumor regions~\cite{deuschel2021multi, shakeri2022fhist}.
Titoriya \textit{et al.} explored few-shot learning to enhance dataset generalization and manageability by utilizing prototypical networks and model agnostic meta learning across four datasets~\cite{titoriya2022few}. The design achieved 85\% accuracy in a 2-way 2-shot 2-query mode~\cite{titoriya2022few}.

% In this paper, we propose a new AI model (DRU-Net) for segmentation of non-small cell lung carcinomas (NSCLCs). 
% Through this work, we have generated a novel dataset of 97 annotated NSCLC WSIs.
% We introduce novelties in three aspects of annotation, model architecture, and augmentation. We annotated samples of two NSCLCs datasets for the first time, created a simple but new cascaded synergistic structure, and proposed a novel spatial augmentation technique called multi-lens distortion and analyzed its effect on model accuracy for various datasets including NSCLCs.
% DRU-Net is a cascaded CNN composed of fused truncated DenseNet201 and ResNet101V2 for patch-wise analysis and a simple lightweight U-Net as the refinement head trained and tested on WSIs of NSCLCs. 
% We also developed adopted models based on the previously proposed state-of-the-art (H2G-Net), a semi-supervised clustering model, and a model trained using few-shot learning~\cite{pedersen2022h2g}. The results of these models are compared and analyzed.
% In addition, saliency maps were utilized in conjunction with the proposed model (DRU-Net) to analyze and elucidate the regions that most significantly impacted the classification of patches, enhancing the model's explainability.
In this paper, we propose a new AI model (DRU-Net) for segmenting non-small cell lung carcinomas (NSCLCs). It is an end-to-end approach consisting of a dual head for feature extraction and patch classification followed by a U-Net for refining the segmentation result. The method is trained and tested on a novel in-house dataset of 97 annotated NSCLC WSIs. To increase model performance, we adopted a few shot learning approach during training and added a multi-lens distortion augmentation technique to WSI images.

\section*{Methods}

\subsection*{Cohorts}

In this study, two different collections of NSCLCs were used; the Norwegian Lung Cancer Biobank (NLCB) cohort and the Bergen cohort~\cite{hatlen2014lung, ramnefjell2017vascular}. The NLCB cohort includes histopathological, cytological, biomarker, and clinical follow-up data from patients with suspected lung cancer diagnosed in Central Norway after 2006~\cite{hatlen2011prolonged}. Both diagnostic tumor biopsies and sections from surgical lung cancer specimen are available. The distribution of histological subtypes in each dataset is listed in Table \ref{tab:nsclc_cases}~\cite{yoh2003tnm, travis20142015}. 

% In this study, 42 NSCLC cases from NLCB with the following number of histological were included: adenocarcinoma (16), squamous cell carcinoma (15), large-cell carcinoma (2), adenosquamous carcinoma (2), undifferentiated carcinoma (1), other neuroendocrine lung cancer (3), and other (3).

The Bergen Cohort comprises 438 surgically treated NSCLC patients diagnosed at Haukeland University Hospital, Bergen, Norway from 1993–2010. In this study, 97 NSCLC cases from the Bergen cohort were included. From both cohorts, 4µm tissue sections were stained with hematoxylin and eosin (HE) and scanned using Olympus BX61VS VS120S5 at x40 magnification. 
%csc

%% explaining the staining process
The sections were deparaffinized, rehydrated in ethanol, and immersed in tap water. Hematoxylin staining was applied and the sections were rinsed in water, and then in ethanol. Sections were then stained with alcoholic eosin. Post-staining, the slides were dehydrated in ethanol, placed in TissueClear, and air-dried~\cite{valla2016molecular}.

\begin{table}[ht!]
    \centering
    \caption{Histological subtypes of non-small cell lung carcinoma cases in the Norwegian lung cancer biobank (NLCB) and the Bergen cohort.}
    \begin{tabular}{@{}l*{1}{p{2.5cm}}*{2}{p{4cm}}@{}}
    \toprule
    Histological subtype & NLCB (n) & Bergen cohort - training and validation (n) & Bergen cohort - test (n) \\
    \midrule
    Adenocarcinoma & 16 & 38 & 7\\
    Squamous cell carcinoma & 15 & 32 & 10\\
    Other non-small cell carcinoma & 11 & 7 & 3\\
    \bottomrule
    Total number of whole slide images & 42 & 77 & 20\\
    \bottomrule
    \end{tabular}
    \label{tab:nsclc_cases}
\end{table} 

To conduct a broader study of the proposed augmentation's effect, we utilized the following open datasets: MNIST, Fashion-MNIST, CIFAR-10, and CIFAR-100~\cite{deng2012mnist, xiao2017fashion, krizhevsky2009learning}.

\subsection*{Ethical aspects}
All methods were carried out in accordance with relevant guidelines and regulations, and the experimental protocols were approved by the Regional Committee for Medical and Health Sciences Research Ethics (REK) Norway (2013/529, 2016/1156 and 257624). Informed consent was obtained from all subjects and/or their legal guardian(s) for NLCB in accordance with REK 2016/1156. For subjects in the Bergen cohort, exempt from consent was ethically approved by REK (2013/529).

\subsection*{Annotations and dataset creations}
We used two annotation approaches on WSIs; whole tumor annotation (WTA) and partial selective annotation (PSA).  In the WTA approach, pathologists marked the tumor outline in 97 WSIs from the Bergen cohort. Of these WSIs, 51 were used for training, 26 were used for validation, and 20 were used for testing. All WSIs with tissue microarray (TMA) holes were placed in the test set to prevent potential biased training. The remaining WSIs were randomly separated into the training, validation, and test sets.

% \subsubsection*{Whole tumor annotation and related labeling method}

To reduce time spent by pathologists in making the WTA annotations, initial annotations were first made in 72 cases using two different AI-based segmentation models, i) the H2G-Net model developed for breast cancer segmentation (n=25) and ii) a customized early-stage clustering model based on the corrected annotations from the H2G-Net model (n=47)~\cite{pedersen2022h2g}. Pathologists then manually adjusted the annotated tumor region using the QuPath software~\cite{bankhead2017qupath}. The remaining 25 cases were annotated manually without AI-based segmentation models. A second pathologist reviewed the annotations, and in case of discrepancy, consensus was reached after discussion. The final annotations were exported as binary masks, serving as ground truth.

% \subsubsection{Partial selective annotation and related labeling method}

In the PSA approach, pathologists marked small regions of interest in 42 WSIs from the NLCB cohort. These WSIs were used for training and validation. The marked areas included parts of the invasive tumor, normal alveolar tissue, stromal tissue, immune cells, areas of necrosis, and other non-tumor tissue such as respiratory epithelium, reactive alveolar tissue, cartilage, blood vessels, glands, lymph nodes, and macrophages. The purpose of marking these regions was to save time spent on manual annotations of the whole tumor regions, and to guide the selection of patches intended for use in the patch-wise model’s training. 

\subsection*{Proposed method}

% \subsubsection*{Overall pipeline}
The pipeline of the proposed model (DRU-Net) has two distinct stages, a patch-wise classification (PWC) stage and a refinement stage. The PWC model was trained on the NLCB cohort using the many-shot method, and the refinement U-Net was trained on a set of down-sampled WSIs from the Bergen cohort. In the PWC stage, the model assigns probabilities to each patch of the WSIs (excluding the glass) indicating whether the patch contains tumor tissue or non-tumor tissue. The classifier’s output provides a preliminary assessment of each patch’s nature, based on local features within the patch. The patches are then stitched together to produce a heatmap of the same size as the down-sampled WSIs.

\begin{figure} [ht!]
    \centering
    \includegraphics[width=0.9\linewidth]{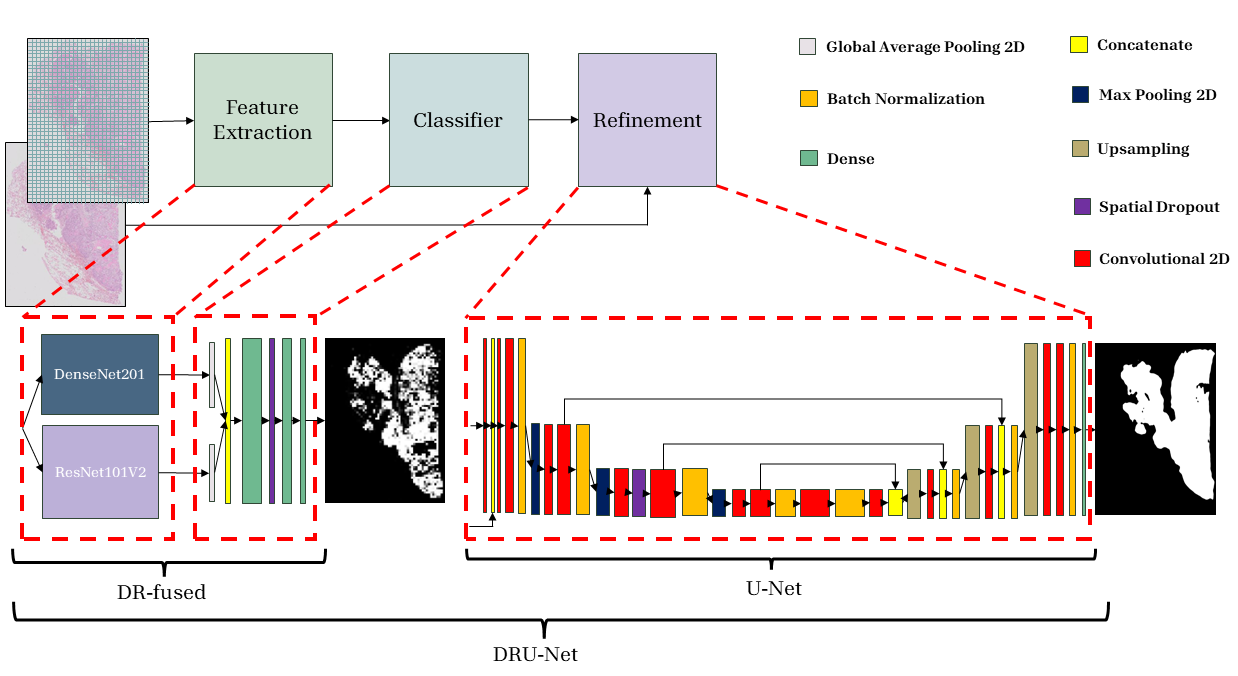}
    \caption{Illustration of the proposed DRU-Net model. The patched image is fed into the classifier part. The output of the classifier is combined with a down-sampled WSI as an input for the refinement head.}
    \label{fig:DRUNet}
\end{figure}

\subsubsection*{Patch-wise classifier}
% The PWC includes a feature extraction and classification step (Figure \ref{fig:DRUNet}). First, features from a given input are generated by two different backbone architectures, DenseNet201 (D) and ResNet101V2 (R), pre-trained on the ImageNet dataset~\cite{deng2009imagenet}. To counter overfitting, both backbones utilize a global average pooling layer to compress the resultant features. Finally, the features from each backbone are then fused before being fed to the classifier head.

The PWC was constructed by fusing truncated backbones of two architectures, DenseNet201 and ResNet101V2, pre-trained on ImageNet~\cite{deng2009imagenet}. These networks are used for parallel processing of the input and feature generation (we refer to this PWC model as DR-fused). In our proposed architecture, both DenseNet201 and ResNet101V2 receive the same input, which is the image patch. Each network processes this input concurrently, and after feature extraction, the outputs from both DenseNet201 and ResNet101V2 pass through their respective global average pooling layers. This step compresses the feature representation and is used to prevent overfitting. The compressed features from both networks are then concatenated and fed through the classifier head (Figure \ref{fig:DRUNet}).

%The training of the PWC was performed using many-shot learning.
%In the development of the PWC of our proposed model, we employed a many-shot training technique.

% These images were captured from the regions of interest annotated in the NLCB dataset, each producing patches in the future steps. Specifically, the 50 images in the non-tumor category included 40 images distinctly lacking tumor characteristics, plus an additional 10 images that exhibited features marginally above the initial threshold, as depicted in Figure \ref{fig:Axis} in supplementary materials. The mentioned threshold was achieved by training the model prior to creating the deliberate imbalance (details explained in the supplementary materials). This deliberate class imbalance was induced after various trials with weighted loss functions, focal loss, threshold tuning, and sampling strategies failed to yield satisfactory generalizability. During the data generation, we subsequently created patches from the mentioned images by randomly cropping a $224\times 224$ pixel section of the image in each epoch, so that each image was used once, providing a different patch in each epoch.

%in every epoch (an epoch refers to one complete cycle through the entire training dataset), ensuring that each image contributed a unique patch per epoch.

\subsubsection*{Refinement network}
 Heatmaps are generated from applying the PWC across the WSIs. The resultant heatmaps are then resized and concatenated with a down-sampled version of the WSI ($1120\times 1120$ pixels). The fused inputs are then fed to a refinement network, similarly, as proposed in H2G-Net~\cite{pedersen2022h2g}. Using a refinement network allows for adjusting the initial patch-wise predictions based on global WSI-level information.

The proposed refinement network is a simple, lightweight U-Net architecture, specifically tailored to process two image inputs (Figure \ref{fig:DRUNet}). In this model, the two inputs (down-sampled RGB WSI and the heatmap) are concatenated into a 4-channel image and then processed through multiple convolutional layers with ReLU activation function. The architecture includes standard components such as Conv2D layers, batch normalization, spatial dropout, skip connections, max pooling for down-sampling, and upsampling layers (nearest-neighbor interpolation). The network concludes with a softmax activation function. 

%To further prevent overfitting, dropout and spatial dropout layers were tested in various positions. The position and dropout type that resulted in the highest DSC was included in the model (Figure \ref{fig:DRUNet}).

\subsubsection*{Data augmentation}
To improve model robustness, data augmentation is commonly performed. Data augmentation generates artificial copies of the training data through some predefined algorithm. This allows the training data to better cover the expected data variation.
Data augmentation was integrated in the data generation process of the training and the following methods were applied randomly: flipping, rotations (multiples of $90^{\circ}$), multiplicative contrast adjustment, hue, and brightness, and the proposed multi-lens distortion augmentation method.
During the many-shot learning using PSA, we extracted patches by randomly cropping a $224\times 224$-pixel section from each image. Each image appeared only once per epoch, where an epoch is defined as one iteration of all the training data.

\subsubsection*{Multi-lens distortion augmentation}
A novel data augmentation method, multi-lens distortion, was developed to simulate several local random lens distortions. This technique aims to allow the model to recognize the important features of the images under a wider range of cell/tissue shapes. 

\begin{figure}[ht!]
    \centering
    \includegraphics[width=0.85\linewidth]{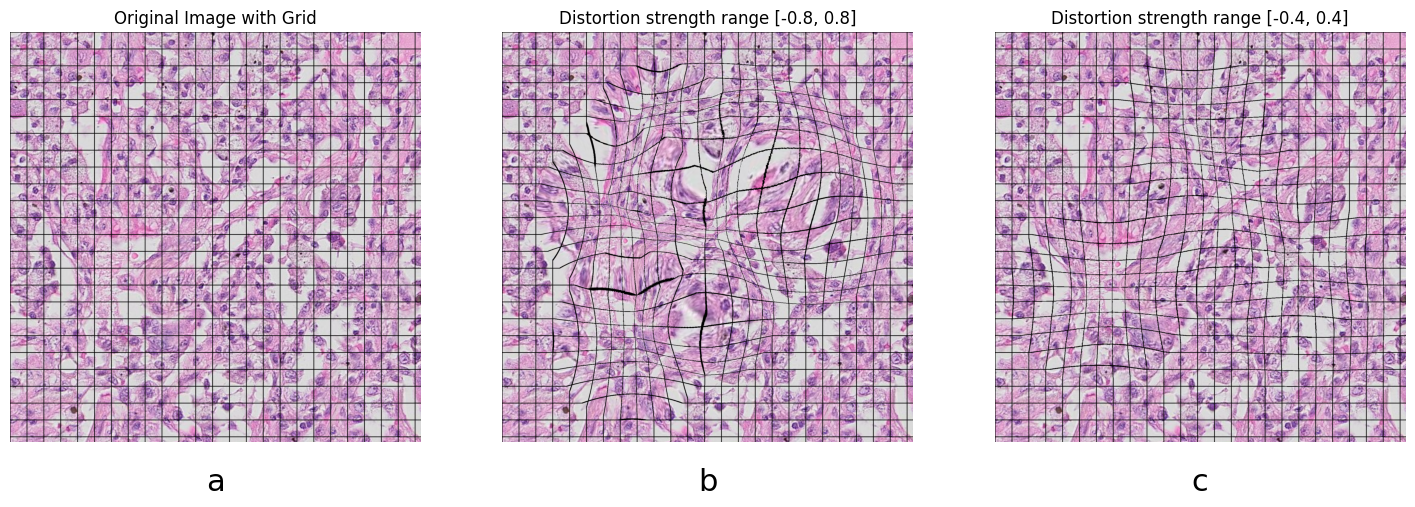}
    \caption{Sample effect of the novel augmentation on a patch with overlaid grids to illustrate the effect. a) Original image showing epithelial cells. b) Augmented image with parameters set too high, cell size variation and deformation are visible. c) Augmented image with a medium setting of the parameters.}
    \label{fig:aug}
\end{figure}

The algorithm uses a fixed number of lenses. For each lens, a random position in the image is selected. A random distortion radius and strength is then used to apply the barrel and/or pincushion distortion effect at the position (Algorithm \ref{alg:multi-lens-distortion}). An example of this augmentation is shown in Figure \ref{fig:aug}. 

%The effect of this augmentation on the training time was measured using the integrated TensorFlow functions by comparing the time with and without the augmentation and the results were averaged on WSIs and compared between the two~\cite{tensorflow2015}.

\begin{algorithm}
\caption{Multi-Lens Distortion Effect on an Image.}
\begin{algorithmic}
\State \textbf{Input:} $image$ - array (Input image of shape \(H\times W\times C\))
\State \textbf{Input:} $num\_lenses$ - int (Number of lenses to apply)
\State \textbf{Input:} $radius\_range$ - tuple \((min\_radius, max\_radius)\)
\State \textbf{Input:} $strength\_range$ - tuple \((min\_strength, max\_strength)\)
\State \textbf{Output:} $distorted\_image$ - array (Image with lens effects applied)

\State $H, W, C \gets \text{shape of } image$
\State $distorted\_image \gets \text{copy of } image$
\State Generate grid of pixel indices $yidx, xidx$ for $image$

\State Generate $num\_lenses$ random centers $(cx, cy)$ within $[image\_size[0] - radius, image\_size[1] - radius]$

\For{$i \gets 0$ \textbf{to} $num\_lenses-1$}
\State $radius \gets$ random value within $radius\_range$
\State $strength \gets$ random value within $strength\_range$
\State Calculate the Euclidean distance $r$ from each pixel to the lens center $(cx[i], cy[i])$
\State Normalize distances: $normalized\_r \gets \frac{r}{radius}$
\State Calculate scaling factor: $scaling\_factor \gets \max(1 - normalized\_r, 0)$
\State Apply distortion to calculate new pixel positions:
\State $distorted\_y \gets (yidx - cy[i]) \cdot (1 - strength \cdot scaling\_factor) + cy[i]$
\State $distorted\_x \gets (xidx - cx[i]) \cdot (1 - strength \cdot scaling\_factor) + cx[i]$
\State Clip $distorted\_y, distorted\_x$ to image bounds
\State Update $distorted\_image$ with pixels from original image at new positions
\EndFor

\State \Return $distorted\_image$
\end{algorithmic}
\label{alg:multi-lens-distortion}
\end{algorithm}

\subsubsection*{Model training}

The PWC network was fine-tuned to adapt to the specific task by freezing the initial layers. The training parameters included: optimizer—Adamax with a learning rate of $1 \times 10^{-4}$; loss function—Categorical Crossentropy; metrics—F\textsubscript{1}-score; batch size—dynamically determined based on the training generator configuration; epochs—up to 200 with early stopping based on validation loss to prevent overfitting.

The refinement network training involved: optimizer—Adam with a learning rate of $1 \times 10^{-4}$; loss function—Dice loss function, optimized for segmentation tasks; metrics—Thresholded Dice score; batch size—2; epochs—up to 300 with early stopping based on validation loss to prevent overfitting; training environment—utilization of GPU and memory growth settings to optimize hardware usage.

In the WTA method, the same set of slides was used for both PWC and segmentation models’ training. From the 97 slides, 77 slides were randomly chosen and divided into training and validation sets by a ratio of 1/3 (51 and 26 slides, respectively), while 20 slides (including those with TMA holes) were used for testing. 
% Slides with TMA holes were intentionally included in the test set to prevent any potential biased training on the TMA holes, the rest were randomly dived into the mentioned quantities between the training, validation, and test sets.

WSIs in the dataset from the Bergen cohort were divided into tiles (patches) and each tile was fed into the neural network along with the non-tumor/tumor label based on the provided annotation. 
To create the annotation labels for patches, non-tumor and tumor tiles were assigned the values 0 and 1, respectively. We first used a threshold on color gradients to separate the tissue from the background glass. Any tile that did not include more than 25\% tissue was disregarded, meaning that all the input tiles contained less than 75\% background glass. Also, a minimum of 5\% tumor area was required for a tile to be classified as tumor; and for the non-tumor regions, only tiles with no tumor were assigned. Tiles with less than 5\% area of tumor were disregarded.

Using the annotated WSI regions in the NLCB dataset, 40 areas were appointed to the tumor class (labeled as 1) and 50 areas to the non-tumor class (labeled as 0).
The selected areas led to the generation of patches in subsequent steps. Specifically, out of 50 areas categorized as non-tumor, 40 clearly lacked tumor characteristics, and 10 showed features slightly above the initially-achieved threshold, as shown in Supplementary Fig. S2. This threshold was established through model training before intentionally creating an imbalance in the dataset. 
The imbalance was introduced after unsuccessful attempts to enhance model generalizability through various methods, including weighted loss functions, focal loss, threshold adjustment, and sampling strategies.

\subsubsection*{Post-processing}
After the segmentation results were received, two post-processing steps were performed. First, small fragments were removed by converting images into grayscale and then to binary format to identify and eliminate fragments smaller than certain size threshold. The threshold was set as the smallest segmentation area annotated in our ground truth.
% In the second step, an edge smoothing algorithm was applied. This was done using morphological operations to reduce jagged edges and make the objects' boundaries smoother. We applied morphological opening (erosion followed by dilation) with a kernel size of $7\times 7$ and a median blur with a kernel size of $11\times 11$ for smoothing edges.
In the second step, an edge smoothing algorithm was applied to enhance image quality. This improvement was achieved through mathematical techniques known as morphological operations, which are commonly used in digital image processing to modify the geometrical structure of images. Specifically, we used a process called morphological opening, which involves an erosion operation followed by a dilation. This sequence helps reduce jagged edges and smooths the boundaries of objects within the image. The operations were performed using a kernel size of $7\times 7$. Additionally, a median blur with a kernel size of $11\times 11$ was applied to further smooth the edges. It’s important to note that these morphological operations are purely computational methods used to process the digital images and should not be confused with the morphological study of biological tissues.

\subsection*{Implementation}
% \subsubsection*{Software and hardware configurations}
% All experiments were conducted on a desktop with Ubuntu Linux 20.04 operating system, an Intel Xeon Gold 6230 @2.10GHz central processing unit (CPU) with 256 GB RAM, and an NVIDIA Quadro RTX 6000 dedicated graphics processing unit (GPU), and a regular solid-state drive. 
Implementation was done in Python 3.8.10. TensorFlow (v2.13.1) was used for model architecture implementation and training~\cite{tensorflow2015}. These additional libraries were used for the experiments: pyFAST, OpenCV, NumPy, Pillow, SciPy, scikit-learn, and Matplotlib~\cite{smistad2015fast, smistad2019high, opencv_library, clark2015pillow, harris2020array, 2020SciPy-NMeth, scikit-learn, Hunter:2007}.
%More information about dependencies and versions can be found at \url{https://github.com/AICAN-Research/DRU-Net}.
% \subsubsection*{Deployment}
Trained models were converted to the ONNX format using the tf2onnx library~\cite{onnx_tensorflowonnx}.
Converted models were then integrated into FastPathology for deployment~\cite{pedersen2021fastpathology}. 
FastPathology is an open-source, user-friendly software developed for deep learning-based digital pathology that offers tools for processing and visualizing WSIs.
The source code used to conduct the experiments is made openly available at \url{https://github.com/AICAN-Research/DRU-Net}.

\subsection*{Experiments}
To compare the proposed model (DRU-Net) with other models, the following experiments were carried out: modifications of the previously introduced H2G-Net model on both datasets, DRU-Net with the backbone trained on the Bergen cohort and NLCB, and applying the few-shot and many-shot learning techniques along with clustering (Table \ref{tab:Experiments})~\cite{pedersen2022h2g}.

% \subsubsection*{H2G-Net and its refinement to lung data}
H2G-Net could be tested as is, and be fine-tuned with five different modifications~\cite{pedersen2022h2g}. First, H2G-Net was tested without any modification, fine-tuning, or additional training, to see whether a model trained for breast cancer tumor delineation can also work for lung cancer. Second, the PWC of the H2G-Net was fine-tuned on annotated NSCLCs from the Bergen cohort, and the original U-Net of H2G-Net was applied on top of the PWC results. Third, the whole model (PWC and U-Net) was fine-tuned on the training data. Then, the same three methods were tested, but with the PWC trained on NLCB instead of the Bergen cohort.

\begin{table}[ht!]
    \centering
    \caption{Methods and experiments carried out with various models on the same 20 WSIs of the test set from the Bergen cohort. Abbreviations: PWC: patch-wise classifier; NLCB: Norwegian Lung Cancer Biobank; FSC: few-shot (with a pre-trained MobileNetV2~\cite{sandler2018mobilenetv2} model) + clustering; MSC: many-shot (with a pre-trained MobileNetV2~\cite{sandler2018mobilenetv2} model) + clustering.}
    \begin{tabular}{rlcc}
        \toprule
        & \textbf{Models} & \textbf{Modifications} & \textbf{Training dataset(s)} \\
        \midrule
        \textbf{(\rom{1})} & H2G-Net & --- & --- \\ 
        \textbf{(\rom{2})} & H2G-Net & Fine-tuned PWC & Bergen Cohort\\ 
        \textbf{(\rom{3})} & H2G-Net & Fine-tuned U-Net & Bergen Cohort\\ 
        \textbf{(\rom{4})} & H2G-Net & Fine-tuned PWC and original U-Net & Bergen Cohort\\
        \textbf{(\rom{5})} & DRU-Net &  --- & Bergen Cohort \\
        \textbf{(\rom{6})} & H2G-Net & Fine-tuned PWC  & NLCB\\ 
        \textbf{(\rom{7})} & H2G-Net & Fine-tuned PWC and U-Net &PWC trained on NLCB, U-Net trained on Bergen Cohort\\ 
        \textbf{(\rom{8})} & FSC & --- &NLCB\\ 
        \textbf{(\rom{9})} & MSC & --- &NLCB\\ 
        \textbf{(\rom{10})} & DRU-Net & --- & PWC trained on NLCB, U-Net trained on Bergen Cohort\\
        \bottomrule
    \end{tabular}
    \label{tab:Experiments}
\end{table}

% \subsection*{Ablation studies}
% \subsubsection*{Evaluation of the proposed augmentation}
An ablation study was performed to evaluate the effect of the proposed multi-lense distortion augmentation. A pre-trained DenseNet121 was tested on four open datasets: MNIST, Fashion-MNIST, CIFAR-10, and CIFAR-100~\cite{deng2012mnist, xiao2017fashion, krizhevsky2009learning}.
Experiments were repeated with and without this augmentation on the mentioned open datasets by randomly selecting 10\% of the training data and the results were compared using Wilcoxon test (Table \ref{tab:multi-lens_impact}). Both control and test groups included other augmentation techniques such as color adjustments, flipping, rotation, brightness, and contrast augmentations.
The effect of this augmentation on the training time was measured using the integrated TensorFlow functions by comparing the time with and without the augmentation and the results were averaged on WSIs and compared between the two~\cite{tensorflow2015}.

% \subsubsection*{Experiments to counter class imbalance}
% % We observed an extreme class imbalance in our patches from the annotated data, particularly within the context of varying tissue types. 
% To solve the low precision resulted from the existing imbalance in data and mixed features from the two classes, a multifaceted approach was employed to enhance model performance. This included resampling, under- and over-sampling, the application of focal loss~\cite{lin2017focal}, sampling from clustered tissue types, weighted loss function, and threshold tuning (see supplementary materials for a detailed explanation of the methods tested). When training the many-shot model, a different deliberate imbalance was induced in the data to adjust for the model's threshold as explained previously in the \textit{Patch-wise classifier} subsection.

% \subsubsection*{Impact of skip connections in refinement}
We also investigated the effect of removing the top-most skip connection of the U-Net refinement model and we calculated the average Hausdorff distances (HDs) for two sets of final segmentation predictions in comparison to a ground truth set. This was done to quantify the effect of removing that skip connection, which was done to reduce the small fragments around the segmentation perimeter.

\subsection*{Model evaluation}

\subsubsection*{Quantitative model assessment}
To quantitatively validate the patch-wise classification performance, precision, recall, and F\textsubscript{1}-score were used~\cite{goutte2005probabilistic}. The validation of the final segmentation on WSI-level was performed using DSC and HD~\cite{kim2015quantitative}. 
% More information about these metrics can be found in the Supplementary information.

\subsubsection*{Qualitative model assessment}
The qualitative assessment of the segmentation results was conducted by two pathologists using the scoring system described in Table \ref{tab:Qualitative}. Qualitative assessment was done on the same 20 WSIs of the test set from the Bergen cohort.

\begin{table}[ht!]
\begin{center}
    \caption{Qualitative evaluation scoring system.}
    \begin{tabularx}{0.9\textwidth}{|>{\raggedright\arraybackslash}X|>{\raggedright\arraybackslash}X|>{\raggedright\arraybackslash}X|>{\raggedright\arraybackslash}X|>{\raggedright\arraybackslash}X|>{\raggedright\arraybackslash}X|} 
     \hline
     0 & 1 & 2 & 3 & 4 & 5\\
     \hline
     No tumor tissue in image or segmentation, or image not suitable for analysis & Completely wrong segmentation of tumor, tumor tissue not segmented & A large part of the tumor is not segmented & Most of the tumor is correctly segmented, but some false positive or false negative areas & Most of the tumor is correctly segmented, only sparse false positive or false negative areas & The whole or almost the whole tumor correctly segmented \\ 
     \hline

\end{tabularx}
\label{tab:Qualitative}
\end{center}
\end{table}

\subsubsection*{Saliency maps}
To survey the model's decision-making process and the areas of patches that were most relevant for predicting the tumor class, we employed a method known as gradient-based saliency maps~\cite{patro2019u, simonyan2013deep, zeiler2014visualizing, sundararajan2017axiomatic}. This approach operates by computing the gradient of the output class (the class for which we want to understand model sensitivity) with respect to the input image. These gradients indicate the sensitivity of the output to each pixel in the input image. By highlighting the pixels with the highest gradients, we can visualize the areas that most strongly influenced the model's classification decision.
We used six different patches for this test selected from six different WSIs from the Bergen cohort. Patches were chosen to represent true and false positive predictions. Patches with true positive predictions were selected to include various histological features and cell types in each patch to better assess the model's decision process.

\section*{Results}

\begin{figure} [ht!]
    \centering
    \includegraphics[width=0.9\linewidth]{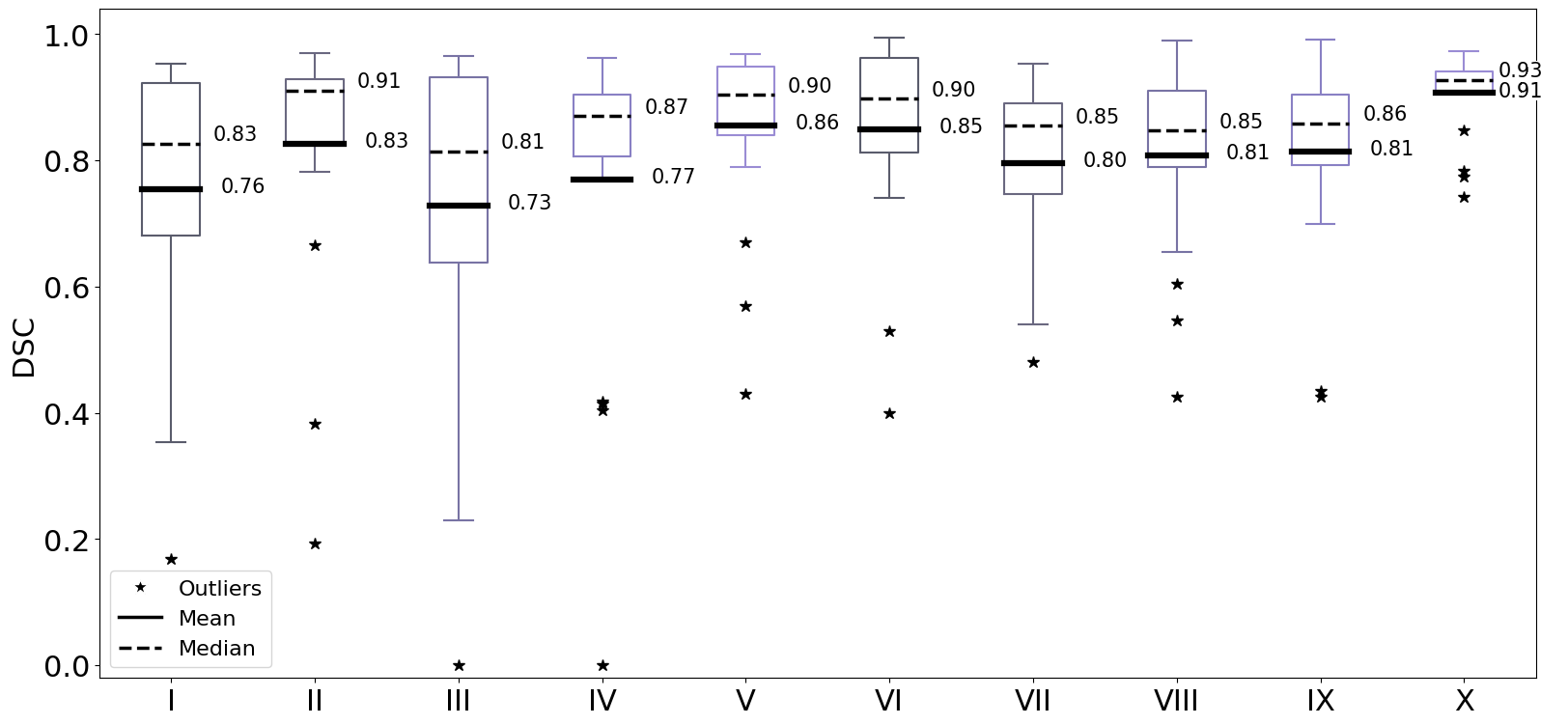}
    \caption{Boxplots of the Dice similarity coefficients (DSCs) of the experiments shown Table \ref{tab:Experiments} on the 20 WSIs of the test set. \textbf{I)} original H2G-Net, \textbf{II)} H2G-Net with fine-tuned PWC on Bergen cohort, \textbf{III)} H2G-Net with fine-tuned U-Net on Bergen cohort, \textbf{IV)} H2G-Net with fine-tuned PWC and U-Net on Bergen cohort, \textbf{V)} DRU-Net trained on Bergen Cohort,  \textbf{VI)} H2G-Net with fine-tuned PWC on NLCB, \textbf{VII)} H2G-Net with fine-tuned PWC on NLCB and fine-tuned U-Net on Bergen Cohort, \textbf{VIII)} FSC, \textbf{IX)} MSC,  \textbf{X)} DRU-Net with PWC trained on NLCB and U-Net trained on Bergen Cohort.}
    \label{fig:DSC_results}
\end{figure}

\begin{figure} [ht!]
    \centering
    \includegraphics[width=0.85\linewidth]{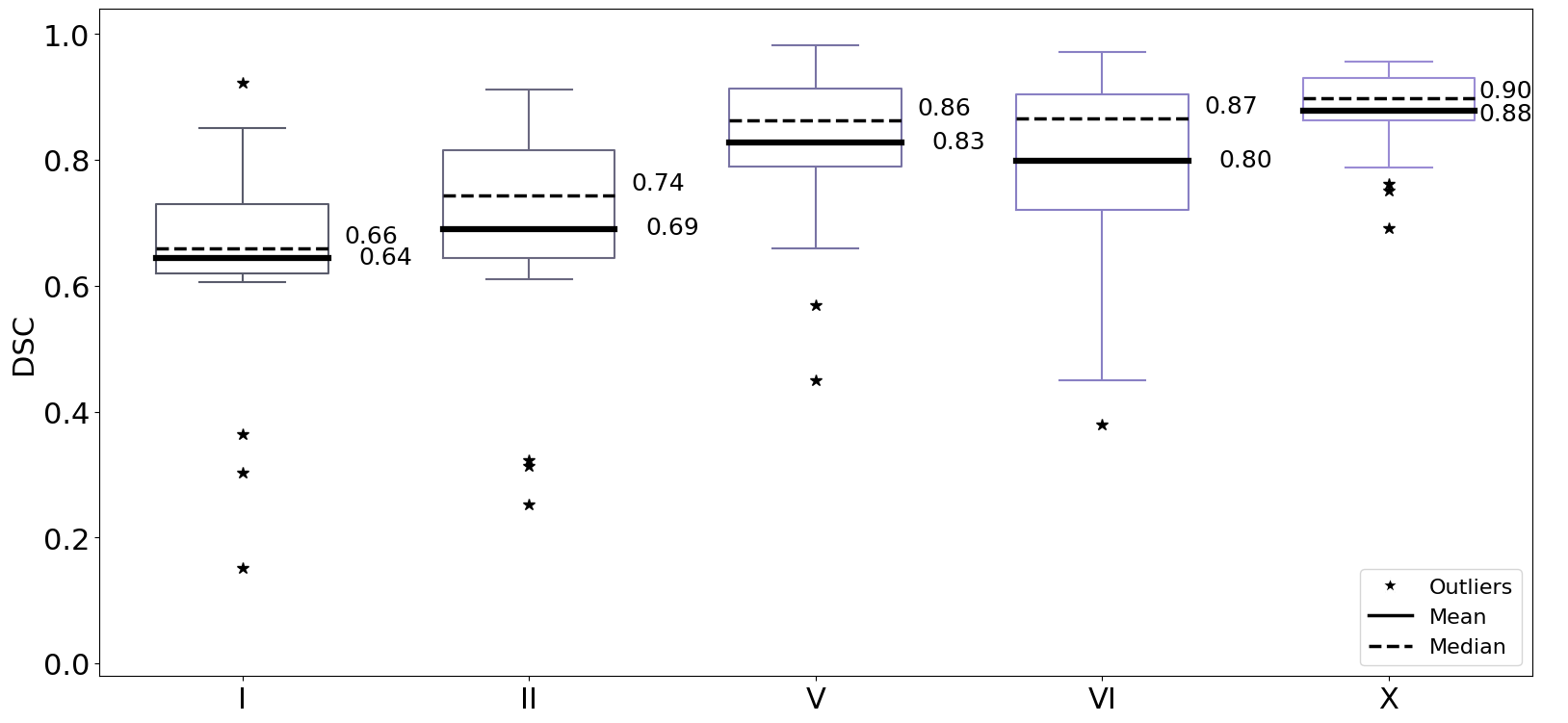}
    \caption{Boxplot of the Dice similarity coefficients (DSCs) of the PWC models in experiments listed in Table \ref{tab:Experiments} without the refinement network, only the patch-wise classifier is used to produce these results. \textbf{I)} original H2G-Net, \textbf{II)} H2G-Net with fine-tuned PWC on Bergen cohort, \textbf{V)} DR-fused trained on Bergen Cohort,  \textbf{VI)} H2G-Net with fine-tuned PWC on NLCB, \textbf{X)} DR-fused trained on NLCB and U-Net trained on Bergen Cohort.}
    \label{fig:PWC_DSC}
\end{figure}

% Figure \ref{fig:DSC_results} summarizes the DSC results for the tested experiments listed in Table \ref{tab:Experiments}. The results of the experiments without the refinement networks are shown in Figure \ref{fig:PWC_DSC}.
Highest DSC in average on the 20 WSIs of the test set from Bergen cohort was achieved by DRU-Net followed by the H2G-Net with fine-tuned PWC on the Bergen cohort (Figure \ref{fig:DSC_results}). Similar differences in DSC were observed for the models without the refinement networks (Figure \ref{fig:PWC_DSC}).

\begin{figure}[ht!]
    \centering
    \includegraphics[width=0.9\linewidth]
    {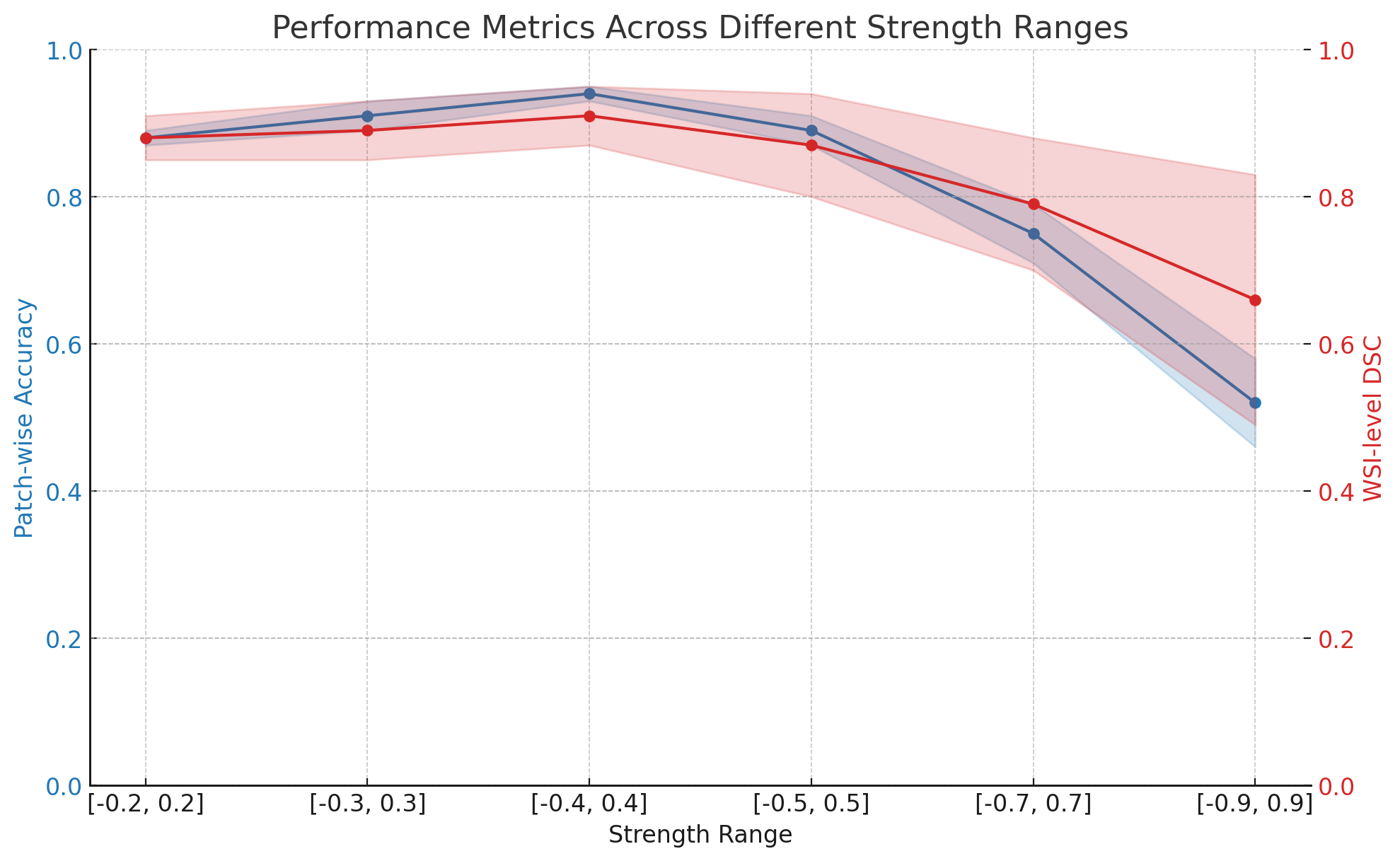}
    \caption{The impact of the multi-lens distortion augmentation technique using the DRU-Net model. DSC: Dice similarity coefficient. The highlighted regions indicate the variance, and the mean values are shown on the curve.}
    \label{fig:AugStats}
\end{figure}

% \subsection*{Novel augmentation results}
% Table \ref{tab:multi-lens_impact} shows the proposed augmentation's effect when applied to public multi-class datasets.
Proposed multi-lens distortion augmentation applied to various datasets resulted in increased F\textsubscript{1}-score overall, this change was statistically significant when applied to our dataset from the NLCB (Table \ref{tab:multi-lens_impact}).
Applying this augmentation technique increased training time by an average of 8\%. 
% Figure \ref{fig:AugStats} shows the effect of the augmentation's strength on our proposed model's performance.
DSC and patch-wice accuracy were increased when the multi-lens distortion augmentation was used with a strength of magnitude in the range [0.2, 0.4], but higher magnitudes caused a decrease in the performance (Figure \ref{fig:AugStats}).

\begin{table}[ht!]
    \centering
    \caption{The impact of the multi-lens distortion augmentation technique using different architectures on different datasets with randomly selecting 10\% of the training data. Pairwise tests were performed using Wilcoxon signed-rank tests.
    % (tests the null hypothesis that two related paired samples come from the same distribution)
    The augmentation design with the highest F\textsubscript{1}-scores row-wise are highlighted in bold.}
    \begin{tabular}{cc|cc|c}
        \toprule
         & & \multicolumn{2}{c}{\textbf{F\textsubscript{1}-score}} & \\ \cmidrule(lr){3-4}
        \textbf{Model} & \textbf{Dataset} & \textbf{wo/aug} & \textbf{w/aug} & \boldsymbol{$p$}\textbf{-value} \\
        \midrule
         DenseNet121 & MNIST & 0.9893 & \textbf{0.9894} & 0.2311 \\ 
         DenseNet121&  Fashion-MNIST& 0.9043 & \textbf{0.9208} & <0.001 \\ 
         DenseNet121&  CIFAR-10 & 0.8086 & \textbf{0.8235} & <0.001 \\ 
         DenseNet121&  CIFAR-100 & 0.5199 & \textbf{0.5581} & 0.0502 \\
         H2G-Net & NLCB & 0.8299 & \textbf{0.8341} & 0.0701 \\
         DRU-Net&  NLCB& 0.8868 & \textbf{0.9025} & 0.0241 \\
         \bottomrule
    \end{tabular}

    \label{tab:multi-lens_impact}
\end{table}

% \subsection*{Baseline H2G-Net on lung slides}
The original H2G-Net resulted in an average of 0.76 DSC (Figure \ref{fig:DSC_results}) and 0.66 intersection over union (IOU) scores. On average, 25\% of the non-tumor regions around the true tumor outlines were falsely labeled as tumor. When the PWC part of the model was used without refinement, the predictions resulted in 0.64 DSC and 0.61 IOU, showing that the refinement improved the predictions significantly.

% \subsection*{Transferred H2G-Net on lung slides}
Fine-tuned PWC trained and validated on 77 WSIs from the Bergen cohort with direct implementation of pre-trained U-Net from H2G-Net was tested on 20 WSIs from the Bergen cohort and resulted in an average of 0.83 DSC (median 0.91) (Figure \ref{fig:DSC_results}) and an average 0.74 IOU scores. Scores were reduced to an average of 0.77 DSC (median of 0.87) and an average of 0.69 IOU when both the U-Net and the PWC were fine-tuned.

\begin{figure}[h!]%{r}{0.4\textwidth} 
    \centering
    \includegraphics[width=1\textwidth]{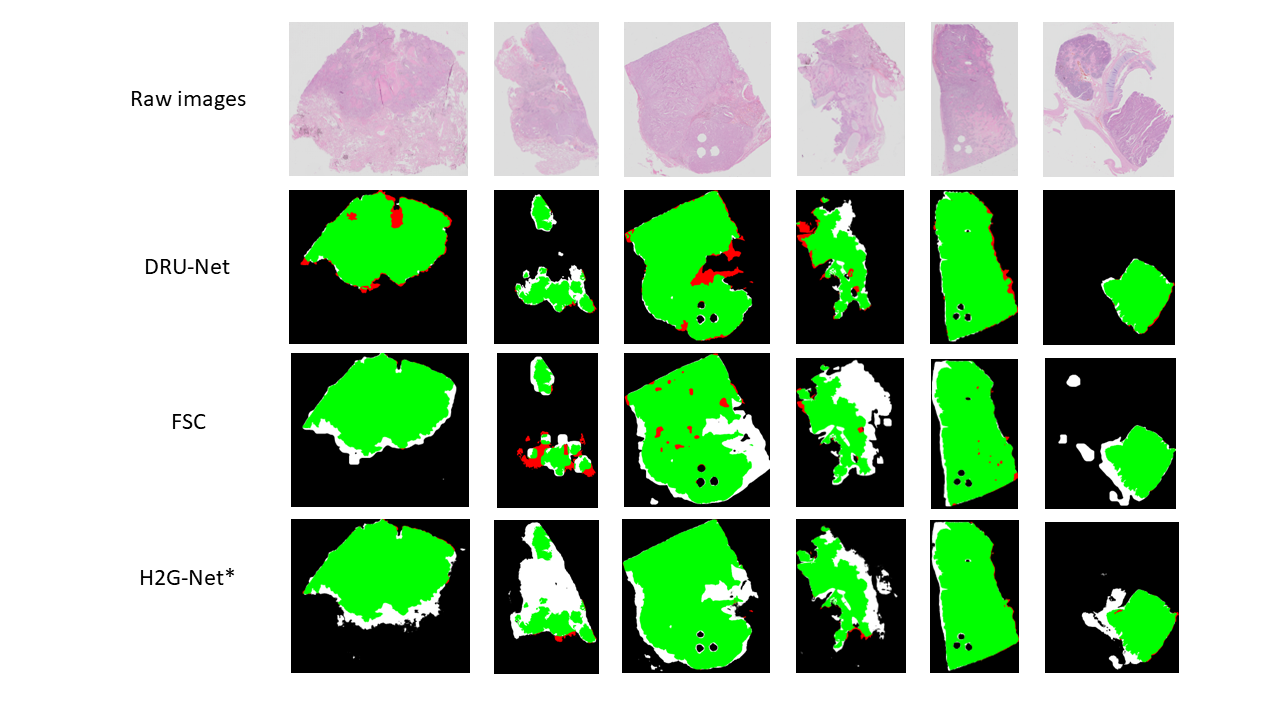}
    \caption{Sample results of three tested networks. First row: original whole slide images (WSIs), second row: DRU-Net, third row: FSC (Few-shot learning + clustering), fourth row: H2G-Net with fine-tuned patch-wise classifier and original U-Net. Green pixels indicate true positives, White pixels indicate false positives and red pixels indicate false negatives.}
    \label{fig:SegResults}
\end{figure}

% \subsection*{Proposed model}
The proposed model (DRU-Net) tested on the same 20 WSIs resulted in an average of 0.91 DSC (median 0.93) and 0.81 IOU. Also removing the top skip connection in our U-Net model (DRU-Net) resulted in the average reduction of HD by 4.8\%.
Figure \ref{fig:SegResults} shows a comparison of the results from various models. Table \ref{tab:Backbones} summarizes various backbones' performance in the patch-wise classifier part of the model.

\begin{table}[ht!]
    \centering
    \caption{Comparison of different backbone architectures for patch-wise classification of lung cancer tissue using the many-shot method. The best-performing architecture per metric is highlighted in bold. Abbreviations: DR: fusion of DenseNet201 (D) and ResNet101V2 (R).}
    \begin{tabular}{r|ccc}
        \toprule
        \textbf{Architecture} & \textbf{F\textsubscript{1}-score} & \textbf{Precision} & \textbf{Recall} \\
        \midrule
        VGG19~\cite{simonyan2014very} & 0.87 & 0.86 & 0.87\\
        ResNet101V2~\cite{he2016deep} & 0.89 & 0.89 & 0.89\\
        MobileNetV2~\cite{sandler2018mobilenetv2} & 0.86 & 0.86 & 0.86\\
        EfficientNetV2~\cite{tan2021efficientnetv2} & 0.89 & 0.89 & 0.89\\
        InceptionV3~\cite{szegedy2016rethinking} & 0.90 & 0.89 & 0.91\\
        DenseNet201~\cite{huang2017densely} & 0.91 & 0.91 & 0.91\\
        Proposed DR-fused & \textbf{0.94} & \textbf{0.94} & \textbf{0.93}\\
        \bottomrule
    \end{tabular}
    \label{tab:Backbones}
\end{table}

% \subsection*{Runtime comparison}
We compared the performance of several models on processing a set of 20 WSIs with average dimensions being approximately 108\,640 pixels in width and 129\,835 pixels in height. H2G-Net and its fine-tuned versions were the fastest models during inference (62 seconds). Although the many-shot and few-shot models had faster training they exhibited slower runtimes, with MSC taking the longest at 167 seconds and DRU-Net at 152 seconds.

%\hfill \break

\begin{figure}[ht!]%{r}{0.9\textwidth} 
    \centering
    \includegraphics[width=0.9\textwidth]{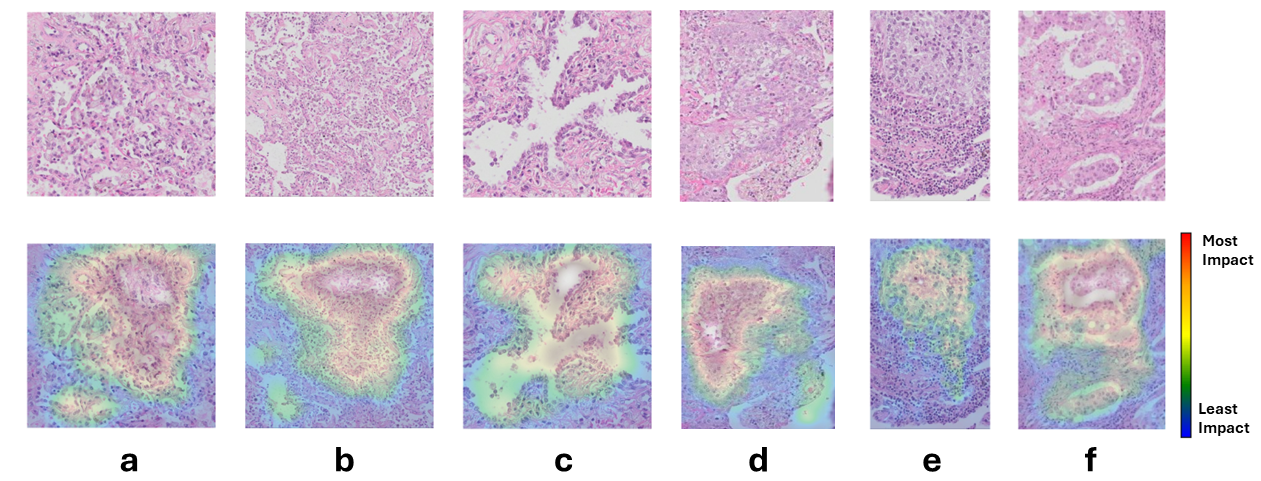}
    \caption{Sample patches (top row) and their overlaid saliency maps (bottom row) where only the patches were given to the PWC model. Note that the saliency map does not indicate malignancy; instead, it shows how different regions of the image influence the classification decision. The colors on the map range from blue, indicating the least influence, to red, which indicates the most influence. Figures \textbf{a}, \textbf{b}, and \textbf{c} show false positive tumor detection. Figures \textbf{d}, \textbf{e}, and \textbf{f} show true positive tumor detection.}
    \label{fig:Saliency}
\end{figure}

Results of saliency map analysis on 6 patches are shown in Figure \ref{fig:Saliency}. 
False positive areas in the saliency maps were partly explained by areas with reactive pneumocytes, macrophages, and reactive pneumocyte hyperplasia.

The qualitative assessment resulted in an average score of 3.95 out of 5. In 9 of the assessed cases, there were sparse areas in the periphery that the model misclassified.

\section*{Discussion}

% TOPIC: Summarize what has been done and most important findings

In this paper, we introduce a novel AI-based model to segment the outline of NSCLCs. We have incorporated a patch-wise classifier, synergistically integrating truncated DenseNet201 and ResNet101V2 architectures, which is enhanced by a segmentation refinement model adopting a streamlined U-Net framework. This composite model demonstrated superior performance over other tested backbones. 
This study also resulted in a novel dataset comprising annotated NSCLCs and marked regions of interest in WSIs from NSCLCs, including different tissue types.
Our results also indicate that the PSA approach yielded more effective training outcomes for the patch-wise classifier than WTA techniques with and without class balancing using tissue clustering. 

% MLD
Our study also demonstrated that the implementation of the multi-lens distortion augmentation technique enhanced classification outcomes across diverse datasets with a limited volume of training data. However, it is important to acknowledge that the effect of this augmentation can vary depending on the data itself. We investigated the effect of the augmentation's strength range on the patch-wise accuracy and refinement network's DSC on WSI-level, concluding that the degree of augmentation applied plays a pivotal role in its impact on the training process. Excessively strong distortion of images could obstruct the model's ability to learn relevant patterns as shown in the impact of the multi-lens distortion augmentation with various strength ranges.

% Comparison with others' works
The RGSB-UNet model features a unique hybrid design that combines residual ghost blocks with switchable normalization and a bottleneck transformer~\cite{zhao2023rgsb}. This design focuses on extracting refined features through its complex structure. However, in our study, we found that simpler and more synergistic designs can also effectively extract reliable features.

The MAMC-Net model improves tumor boundary detection by using a conditional random field layer~\cite{zeng2023mamc}, whereas the DRU-Net model enhances segmentation by fine-tuning a U-Net on a down-sampled image. Both methods achieved good results, but our approach using a U-Net on down-sampled images is faster and still highly efficient. Compared to the MAMC-Net study, our PSA approach used a much smaller dataset and achieved similar results.

Similar to H2G-Net, our proposed model, DRU-Net, also utilizes a cascaded design with the two stages of PWC and refinement and has achieved comparable results~\cite{pedersen2022h2g}. Although H2G-Net uses a lightweight PWC and a relatively heavier U-Net for refinement, DRU-Net performed better with a heavier PWC and a lightweight U-Net. This could be due to less training data available in our case which can require a more complex feature extraction process. Pedersen \textit{et al.} also introduced a balancing technique to ensure a balanced representation of the available categories, which helps minimizing bias toward any specific tissue type or tumor characteristic. In our study, we have also used a clustering-based balancing method to reduce bias during sampling from our WTA. However, we have also benefited from our PSA approach with the induced imbalance in generated samples to minimize the bias toward tumor labels, which proved to work better in our test dataset. 

% U-Net from H2GNet and DRU-Net
The decrease in performance after fine-tuning the U-Net layers of the H2G-Net may be due to the relatively small number of annotated WSIs available in our study. Conversely, the DRU-Net network's superior performance under similar conditions suggests the efficacy of the DR-fused network accompanied by a relatively lightweight U-Net architecture in data-scarce scenarios.

% Why H2GNet is not working well here
The relatively low performance of the original H2G-Net on NSCLCs can be explained by different tissue morphology, growth pattern, and stromal invasions, which can cause misleading during inference~\cite{menon2022exploring, kashima2019molecular,petersen2011morphological,inamura2017lung,zhao2023single,binder2021morphological, tan20202019, travis20142015}.

% Existing imbalance challenge
In this study, we encountered challenges due to significant class imbalance between the patches derived from the WTA approach. Addressing the resultant low precision, a comprehensive strategy was implemented to improve model accuracy. Key interventions included resampling techniques, both under- and over-sampling, as well as the incorporation of focal loss, which specifically helps to address class imbalance by modulating the loss function to focus on harder-to-classify examples~\cite{lin2017focal}. Furthermore, we explored the clustering of similar tissue types before sampling, the use of a weighted loss function, and adjustments to the decision threshold.

% Induced imbalance
Additionally, in the training phase of the many-shot model using the samples derived from PSA approach, to maximize the model’s performance, we deliberately introduced a controlled imbalance which was aimed at optimizing the threshold settings.
The deliberate construction of an imbalanced dataset resulted in improved performance. This approach outperformed resampling, under- and over-sampling, focal loss, sampling from clustered tissue types, weighted loss function, and threshold tuning~\cite{lin2017focal}. However, the induced class imbalance, while promising, carries the risk of significant bias. It necessitates careful calibration and continuous monitoring to ensure that the model does not disproportionately favor certain classes or features, leading to skewed results. In our case, the performance was ascertained by testing on an external dataset (the PWC trained using the many-shot method was trained on the NLCB dataset and tested on the 20 slides that were selected for testing from the Bergen cohort).

% Breast tumors originate from epithelial cells in ducts or lobules~\cite{harris2012diseases, henry2020cancer}, showing various cytoplasm characteristics and sometimes abnormal mucin production~\cite{tan20202019, mukhopadhyay2011mucins,marrazzo2020mucinous,astashchanka2019mucin,dreyer2022role}. They often invade surrounding tissue in a stellate or linear manner~\cite{tan20202019}, with potential \textit{in situ} components, marking invasive carcinomas by stromal invasion and desmoplastic reactions~\cite{tan20202019}. Lung tumors, however, arise from air passage epithelial cells, frequently presenting glandular differentiation and possible mucin production~\cite{travis20142015}. Early stages might show less apparent stromal invasion~\cite{travis20142015}, but advanced tumors invade lung parenchyma and other thoracic structures~\cite{travis20142015}. The mentioned differences between the two might explain the large false positive areas in lung datasets when using H2G-Net directly due to these distinct invasion patterns.

% PWC and U-Net affect
Our results indicate that refining the PWC heatmap with the suggested refinement networks improves the accuracy of the tested models. However, the main strengths and weaknesses of the models compared to each other directly stem from the training method used for the PWC models. Additionally, combining the two processes seems to improve and reduce the variance in the segmentation DSC values, indicating that the refinement models have learned to understand overall patterns and connections leading to a better segmentation.

% PSA seems better than WTA
The difference observed in the average DSCs between PWC models indicates that PSA can outperform WTA approaches when the amount of data is relatively limited. 
This is because of the inadequate separability of the features distributions between tumor and non-tumor. 
In the WTA approach, the method involved annotating entire tumor regions, which often included patches where the feature distributions of tumor and non-tumor tissues overlapped significantly. This overlap resulted in inadequate separability, thereby reducing the discriminatory power of the classification models trained using this approach. Consequently, the distinction between tumor and non-tumor features in these patches became less pronounced, leading to potential misclassifications.

Conversely, the PSA method adopted a more selective approach by targeting patches for annotation based on their discriminative morphology. By focusing on patches where the features of each class (tumor and non-tumor) were distinctly separable, PSA enhanced the model's ability to accurately classify these features. This selective annotation process, effectively increased the inter-class variance while reducing the intra-class variance, thus significantly improving the overall performance of the classification models in distinguishing between tumor and non-tumor tissues under conditions of limited data.
In the WTA approach, the mentioned inseparable feature distribution affected the loss function negatively resulting in lower accuracy. This was most likely rooted in the fact that the tumor regions also include other cell types than the invasive epithelial cells. By using histopathological knowledge for selecting areas with the most relevant features in PSA, variation of the features between the two classes could be increased.

% Few-shot + Clustering
Additionally, the study presents evidence that employing few-shot learning in conjunction with a clustering approach can achieve accuracy levels comparable to those reliant on extensive datasets, thereby potentially mitigating the need for large-scale data collection.
The few-shot learning approach is preferable when there is a high degree of similarity within each class of tissue types and a clear distinction between the classes in the feature space~\cite{qi2022task}. 
% This was achieved in our dataset when forming 90 classes of image sets (see Supplementary information for more details).

% Novelty in our Few Shot Learning
In our few-shot learning method, we introduced a novelty by utilizing an evolutionary optimization technique to determine the optimal number of clusters (classes) to minimize intra-cluster variance and maximize inter-cluster variance prior to training. This method ensures that clusters are optimally configured to reflect the most coherent and meaningful class structures, which is crucial when the available training data is scarce. By focusing on minimizing intra-cluster variance and minimizing inter-cluster similarity, the approach enhances the model's ability to generalize from limited examples, a critical aspect in few-shot scenarios where the risk of overfitting is high. Furthermore, evolutionary algorithms offer robust adaptability and flexibility, enabling the model to effectively handle different types and distributions of data. This pre-training optimization led to more efficient training and improved model performance by grouping patches into different classes.

% explainable AI
% For the tumor class, the saliency maps highlight areas within the patches that are characteristic of neoplastic tissue. The consistent emphasis on these areas across various tumor samples suggests that the model has learned to recognize and prioritize these neoplastic features in its decision-making process. They include the size, shape, density, and arrangement of the nuclei, which are well-established criteria in histopathology for identifying tumor cells. The model's focus on these features indicates its ability to learn and apply domain-specific knowledge. This focus can also be observed in the features that lead to false positive results (Figure \ref{fig:Saliency}).

% Qualitative assessments
Qualitative assessment of our results suggests that the DRU-Net model shows limitations in accurately delineating the tumor periphery. This challenge was particularly evident in regions with fibrosis, reactive tissue, or inflammation, where the model tends to produce false positive and false negative segmentations.

% TOPIC: future work
In future work, we suggest reducing the model size, incorporating advanced attention-focusing mechanisms, and using a multi-scale patch-wise classifier to better incorporate information at different scales. 
% Also, there are a few challenges that can be improved regarding the false positives and false negatives. 
% The proposed model requires improvement to correctly delineate the tumor border and differentiating reactive tissues.
% We suggest future studies for gathering more data on areas that are prone to producing false positives or false negatives.
Employing anomaly detection algorithms might help identify reactive tissue outliers that contribute to false positive classifications. 
Additionally, Mask R-CNN architectures are highly effective in distinguishing complex patterns that can be used for a better tumor border delineation. Also, implementing Bayesian neural networks can help in predicting tumor boundaries while quantifying the uncertainty of the predictions. To better use the global information, one can also integrate Markov random fields or conditional random fields along with PWC or transformer architectures to help ensure that the segmented areas are not only based on local pixel values.
We also suggest Neuro-Fuzzy Systems, which leverage the learning capabilities of neural networks with the reasoning capabilities of fuzzy logic to improve the differentiation between the two classes.
To overcome the limited data problem, we can suggest using unsupervised domain adaptation algorithms to leverage annotated data from other histopathology source domains.
% \hfill \break

\section*{Conclusion}
% In conclusion, we have introduced DRU-Net for lung cancer tumor delineation in WSIs. Key points in our approach, which markedly improved classification outcomes, include a fusion of the truncated DenseNet201 and ResNet101V2 networks, deliberately induced class imbalance, manual selection of regions of interest for patch generation, and a multi-lens distortion augmentation. Qualitative assessments revealed a limitation in the model’s ability to precisely delineate the tumor periphery, especially in regions affected by fibrosis, inflammation and reactive tissue.

In conclusion, we have introduced DRU-Net for non-small cell lung cancer tumor delineation in WSIs. Our novel AI-based model, which synergistically integrates truncated DenseNet201 and ResNet101V2 with a U-Net based refinement framework, has demonstrated high performance in NSCLCs over various tested methods. The effectiveness of our patch-wise classifier has been significantly enhanced through the incorporation of the multi-lens distortion augmentation technique and PSA strategy. Additionally, employing few-shot learning with optimized clustering could mitigate the need for large datasets.

\bibliography{main}

\section*{Acknowledgements}

We extend our gratitude to Borgny Ytterhus for her contributions to this project.

\section*{Author contributions statement}

SO created the proposed model and augmentation, conducted the experiments, and wrote the main text. 
MV prepared the data, checked and confirmed the annotations, conducted the qualitative assessments, and helped with the histopathological aspects of the experiments. 
AP assisted with the technical and programming aspects of the work. 
ES provided guidelines for programming and technical aspects of the work. 
SGFD collected and quality-checked all histopathology samples from the NLCB archive. 
VGD annotated the slides, helped with the histopathological aspects of the experiments, and conducted the qualitative assessments. 
MH reviewed and provided feedback on the paper. 
MDH annotated the slides and helped with the histopathological aspects of the experiments. 
TL Provided general guidelines for the experiments and methods. 
MPR contributed in the preparation of the Bergen cohort, reviewed and provided feedback on the paper. 
LAA contributed in the preparation of the Bergen cohort, reviewed and provided feedback on the paper. 
GK provided guidelines for experiments and technical aspects of the work. 
HS was project leader, provided funding and supervision over the whole work and guidelines for the clinical aspects of the work. 
All authors reviewed, revised, and approved the manuscript.

\section*{Data availability}
The datasets generated and/or analysed during the current study are not publicly available due to the sensitive nature of personal medical data from patients who may still be alive.

\section*{Conflict of Interest}

There is no conflict of interest to report.

%The corresponding author is responsible for submitting a \href{http://www.nature.com/srep/policies/index.html#competing}{competing interests statement} on behalf of all authors of the paper. This statement must be included in the submitted article file.

\end{document}